%% file: lb_prp.tex
\documentstyle[preprint,eqsecnum,epsf,aps,floats,tighten]{revtex}  
\input psfig
\input epsf
\input BoxedEPS.tex
\input labelfig.tex

\SetRokickiEPSFSpecial
\HideDisplacementBoxes

\input{definitions}

\begin{document}
\setlength{\unitlength}{1.0mm}
\lefthyphenmin=2
\righthyphenmin=3

\hyphenation{Ari-zo-na} 
\hyphenation{in-clu-sive} 
\hyphenation{lead-ing} 
\preprint{FERMILAB-Pub-00/271-E}
\title{	Inclusive Jet Production in $\bbox{p\bar{p}}$ Collisions }
\input{list_of_authors.tex}
\date{November 10, 2000}
\maketitle
\begin{abstract}
  We report a new measurement of the pseudorapidity (\peta) and transverse-energy 
  (\et) dependence of the inclusive jet production cross section in \ppbar\ 
  collisions at $\sqrt{s}=1.8$~TeV using 95 \ipb\ of data collected with
  the \dzero\ detector at the Fermilab Tevatron. The differential 
  cross section \mycs\ is presented up to $\aeta=3$, significantly extending 
  previous measurements. 
  The results are in good overall agreement with next-to-leading order 
  predictions from QCD and indicate a preference for certain parton 
  distribution functions.
\end{abstract}


\clearpage

This past decade has witnessed impressive progress in both the theoretical 
and experimental understanding of the collimated streams of particles or ``jets'' 
that emerge from inelastic hadron collisions.
Theoretically, jet production in hadron collisions is understood within 
the framework of Quantum Chromodynamics (QCD), as a hard scattering of the 
constituent partons (quarks and gluons) that, having undergone a collision, 
manifest themselves as jets in the final state.
QCD predicts the amplitudes for the hard scattering of partons at high energies.
Perturbative QCD calculations of jet cross 
sections~\cite{EKStheory,Aversatheory,GGKtheory}, using 
accurately 
determined parton distribution functions (PDFs)~\cite{CTEQ,MRST}, have 
increased the interest in jet measurements at the $\sqrt{s}=1.8$ TeV Tevatron 
proton-antiproton collider.
Consequently, the two Tevatron experiments, \dzero\ and CDF,
have served as prominent arenas for studying hadronic jets.

\begin{figure}[!h] \centering
  \begin{picture}(85.5,74)  

  \put(-21.5,-41){\begin{picture}(85.5,73)
	      \epsfysize=16.0cm 
	      \epsfbox{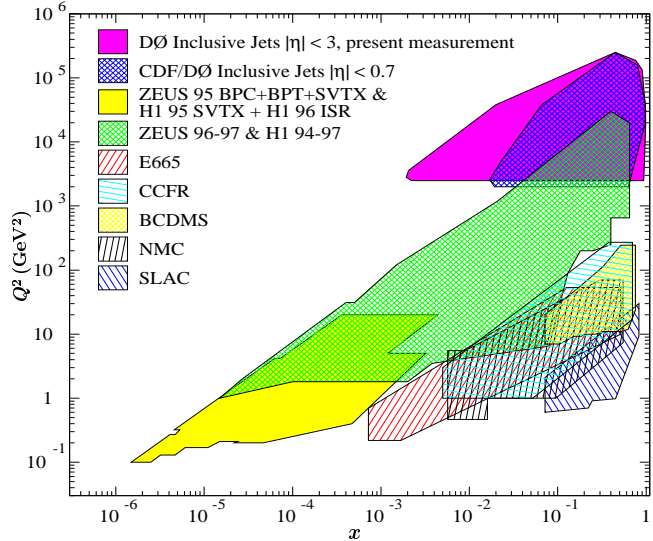}
	    \end{picture}}

  \end{picture}
  \caption{The kinematic reach of this measurement along with that of 
	   other collider and fixed-target experiments in the plane of 
	   the parton momentum fraction $x$ and the square of the 
	   momentum transfer $Q^{2}$.}
  \label{fig:xq2}
\end{figure}

In this Letter, we report a new measurement of the pseudorapidity (\peta)
and transverse-energy (\et) dependence of the inclusive jet production cross 
section~\cite{mythesis}, which examines the short-range behavior of QCD, the 
structure of the proton in terms of PDFs, and possible substructure of quarks 
and gluons.
We present the differential cross section \mycs\ as a function of jet \et\ in 
five intervals of \peta, up to $\aeta=3$, where the pseudorapidity is defined 
as $\eta=\ln \left[ \cot \left( \theta/2 \right) \right]$, with $\theta$ 
being the polar angle.
The present measurement is based on 95~\ipb\ of data collected with the 
\dzero\ detector~\cite{D0detector} during 1994--1995, and significantly 
extends previous measurements~\cite{PDG}, as indicated by the
kinematic reach shown in Fig.~\ref{fig:xq2}.

The primary tool used for jet detection is the compensating, 
finely segmented, liquid-argon/uranium calorimeter, which
provides nearly full solid-angle coverage ($\aeta < 4.1$).
Jets are defined and reconstructed off-line using an iterative fixed-cone
algorithm with a cone radius of $\Rmathcal = 0.7$ in the $\eta$--$\varphi$ 
space, where $\varphi$ is the azimuth.
The missing transverse energy (\met) is calculated from a vector sum
of the individual \et\ values in all the cells of the calorimeter.
Calorimeter cells can occasionally provide spurious noise signals; to 
diminish their effect on jets, such cells are identified and suppressed 
using specific on-line and off-line algorithms. 
 
During data taking, events were selected with a multi-stage trigger system.
The first stage signaled an inelastic \ppbar\ collision.
In the next stage, the trigger required a jet in a calorimeter 
region of $\Delta\eta\times\Delta\varphi=0.8\times1.6$, with \et\ above a 
preset threshold.
In the last trigger stage, selected events were digitized and sent to an 
array of processors.
Jet candidates were reconstructed using a cone algorithm, and the entire
event was recorded if any jet \et\ exceeded a specified threshold.
The four software filters used in this analysis had \et\ thresholds
of 30, 50, 85, and 115 GeV, and accumulated integrated luminosities of
0.364, 4.84, 56.5, and 94.9 \ipb\ respectively~\cite{jetsPRD}.
To present the full range of the data, the cross sections obtained from 
the four jet filters are combined in 
contiguous regions of \et\ in such a way that the more restrictive trigger 
is adopted as soon as it is more than 99\% efficient. 

The position of the primary interaction vertex is reconstructed using
data from the central tracking system.
The two vertices with the largest number of associated tracks are 
retained for further analysis.
At high instantaneous luminosities, multiple interactions are common,
and to correct for inefficiency of the tracking system in identifying 
the primary vertex, we use the global event quantity 
$S_{T}=|\sum\vec{E}_{T}^{\,\mathrm{jet}}|$.
The vertex with the smaller value of $S_{T}$ is defined as the correct event 
vertex, and all kinematic variables are calculated with respect to 
it. 
The dependence of jet \et\ on luminosity was studied, and found 
to be negligible. 
At high pseudorapidities, the jet reconstruction algorithm introduces a 
bias towards $\peta=0$.
Furthermore, the Snowmass jet reconstruction algorithm~\cite{snowmass}
used in the theoretical predictions has a different definition for jet 
angles than that used in the standard \dzero\ off-line algorithm.
Jet \peta\ values are corrected for this difference, which also removes 
any instrumental bias in reconstruction of jet polar angles~\cite{mythesis}.

Backgrounds introduced by electrons, photons, detector noise, accelerator 
losses, or cosmic rays are removed using quality criteria developed for jets
with $\aeta\leq3$. 
To preserve the pseudo-projective nature of the \dzero\ calorimeters, the 
longitudinal ($z$) position of the interaction vertex is required to be within 
50 cm of the detector center; this requirement is $(88.7\pm0.1)\%$ efficient.
A cutoff on \met\ removes background from cosmic ray showers and misvertexed 
events.
\met\ must be smaller than the lesser of $30$ GeV or $0.3\et$ of the leading 
jet if the leading jet is central ($\aeta<0.7$), or less than $0.7\et$ otherwise.
This criterion is nearly $100\%$ efficient. 
Jet quality is based on the pattern of energy deposition in the calorimeter.
The combined efficiency for jet quality ranges from about 99.5\% at lowest 
\et\ and \aeta\ to approximately 98\% at highest \et\ and \aeta.

The jet energy calibration, applied on a jet by jet basis, corrects (on average) 
the reconstructed \et\ for variation in the hadronic response of the calorimeter, 
for the energy associated with underlying spectator interactions, for multiple 
\ppbar\ interactions in the same crossing, noise originating from uranium decay, 
the fraction of any particle's energy that showers outside of the reconstruction 
cone, and for detector non-uniformities.
A complete discussion of the jet energy calibration 
can be found in Ref.~\cite{d0jes}.
An independent test of the jet energy scale, based on the balance in transverse 
energy in 
photon-jet and jet-jet data, confirms the validity of the \dzero\ jet-calibration 
procedure up to $\aeta=3$~\cite{mythesis}. 

In each bin of \peta--\et, the average differential cross section, \mycs, is 
calculated as 
$N/\!\left(\Delta\peta\Delta\et\,\epsilon\!\int\!\!{\mathcal{L}}dt\,\right)$,
where $\Delta\peta\Delta\et$ is the \peta--\et\ bin size, $N$ is the number of 
jets observed in a bin, $\epsilon$ is the total overall efficiency for jet and 
event selection, and $\int\!\!{\mathcal{L}}dt$ represents the integrated luminosity 
of the data sample.
Statistical uncertainties in the values of the cross sections are defined by 
one standard deviation Poisson fluctuations in the associated $N$.

Energy resolution of the \dzero\ calorimeters distorts the jet cross section in \et.
Although the resolution is essentially Gaussian, the jet cross section is shifted 
to larger \et\ due to the steeply falling dependence of jet production on \et.
This effect is removed from the data through an unfolding procedure.
We measure the fractional jet energy resolutions based on the ``same side''
($\eta_{1}\!\!\cdot\!\eta_{2}\!>\!0$) subset of dijet events in the data 
sample.
Using the imbalance in the \et\ of the two leading jets, in each interval of 
\aeta, we parameterize the fractional jet energy resolution as a function of 
jet \et, following the standard description of single-particle energy resolution, 
based on the noise, sampling, and constant terms. 
To determine the amount of distortion in the cross section in each of the five 
\aeta\ intervals, we take an ansatz function of the form 
$e^{\alpha}E_{T}^{\beta}\left( 1 + \gamma 2\et/\sqrt{s} \right)^{\delta}$,
numerically smear it according to the parameterized resolution  in each
\et\ bin, and fit this smeared hypothesis to the observed cross sections
to extract the five sets of four free parameters, $\alpha$, $\beta$, 
$\gamma$, and $\delta$.
The bin-by-bin ratio of the original over the smeared ansatz for each
range of \aeta\ gives the unfolding correction with which we rescale the 
observed cross section to remove the distortion from jet energy 
resolution~\cite{mythesis}.

The jet angular resolution is very good at all \peta, and its effect on the 
cross section is negligible, but it is possible to distort the jet polar angle 
through a mismeasurement of the \z\ position of the vertex.
However, a Monte Carlo study demonstrates that such effects are negligible because 
distortions in jet \et\ are nearly fully compensated by bin-to-bin migrations 
in \aeta\ from the smearing of the \z\ coordinate of the vertex~\cite{mythesis}.

\begin{figure}[!h] \centering
  \begin{picture}(85.5,81)  

  \put(-22.5,-40){\begin{picture}(85,79)
	      \epsfxsize=12.5cm 
	      \epsfbox{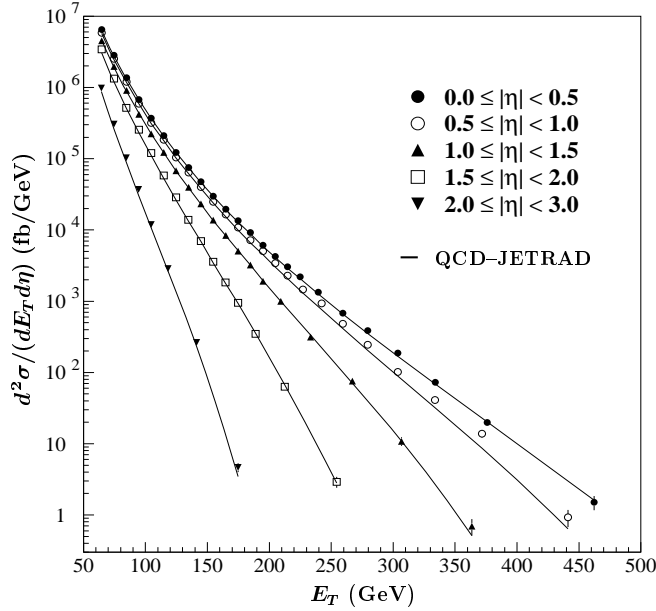}
	    \end{picture}}

  \end{picture}
  \caption{The single inclusive jet production cross section 
	   as a function of jet \et, in five pseudorapidity intervals, showing 
	   only statistical uncertainties, along with theoretical predictions.}	 
  \label{fig:rapdep}
\end{figure}

The final measurements in each of the five \aeta\ regions, along with 
statistical uncertainties, are presented in Fig.~\ref{fig:rapdep} (tables 
of the measured cross sections can be found in Refs.~\cite{mythesis,data_on_www}).
The measurement spans about seven orders of magnitude and extends to the 
highest jet energies ever reached.
Figure~\ref{fig:rapdep} also shows ${\mathcal{O}}\!\left(\alpha_{s}^{3}\right)$ 
theoretical predictions from {\sc JETRAD}~\cite{GGKtheory} with renormalization 
and factorization scales set to half of the \et\ of the leading jet and using the 
CTEQ4M PDF.

\begin{figure}[!t] \centering
  \begin{picture}(85.5,97)  

  \put(-21,-29.3){\begin{picture}(85,97)
	      \epsfysize=15.90cm 
	      \epsfbox{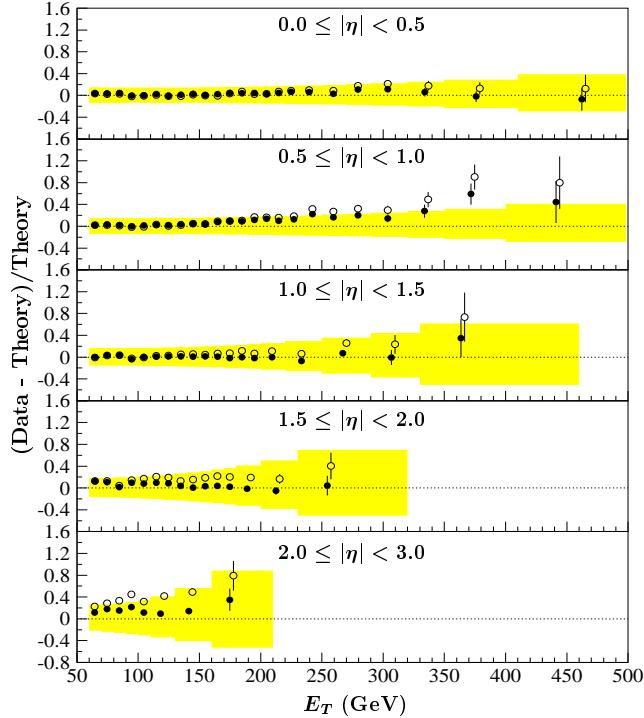}
	    \end{picture}}

  \end{picture}

  \caption{Comparisons between the \dzero\ single inclusive jet cross sections 
           and the ${\mathcal{O}}\!\left(\alpha_{s}^{3}\right)$ QCD predictions 
	   calculated by {\sc JETRAD}  
	   with the CTEQ4HJ ({\boldmath$\bullet$}) and CTEQ4M ({\boldmath$\circ$}) 
	   PDFs. The highest \et\ points are offset slightly for CTEQ4M.}
  \label{fig:comp1}
\end{figure}

\begin{figure}[!t] \centering
  \begin{picture}(85.5,97)  

  \put(-21,-29.3){\begin{picture}(85,97)
	      \epsfysize=15.90cm 
	      \epsfbox{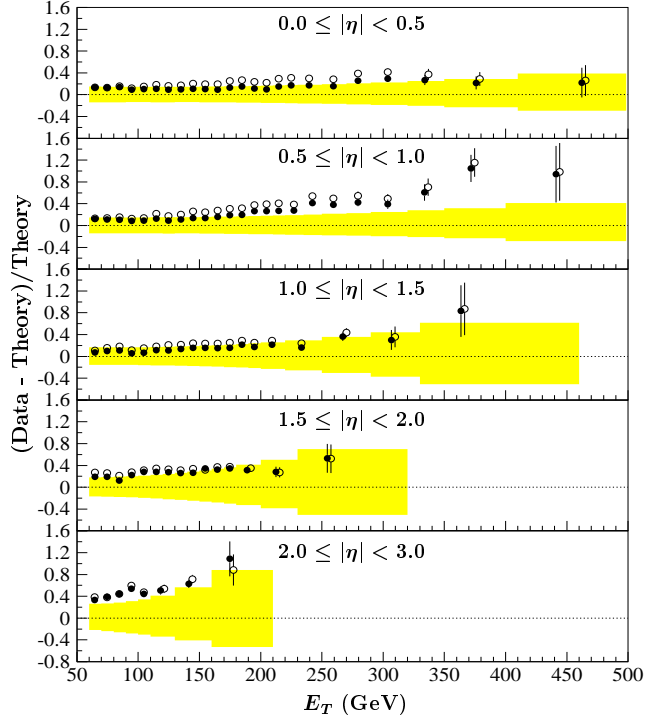}
	    \end{picture}}

  \end{picture}

  \caption{Comparisons between the \dzero\ single inclusive jet cross sections 
           and the ${\mathcal{O}}\!\left(\alpha_{s}^{3}\right)$ QCD predictions 
	   calculated by {\sc JETRAD}  
	   with the MRSTg$\uparrow$ ({\boldmath$\bullet$}) and MRST ({\boldmath$\circ$}) 
	   PDFs. The highest \et\ points are offset slightly for MRST.}
  \label{fig:comp2}
\end{figure}

Figures~\ref{fig:comp1} and~\ref{fig:comp2} provide more detailed comparisons 
to predictions on a linear scale for several PDFs (for other PDFs, see Ref.~\cite{mythesis}).
The error bars are statistical, while the shaded bands indicate one
standard deviation systematic uncertainties.
Because the theoretical uncertainties due to variations in input parameters 
are comparable to the systematic uncertainties~\cite{errorstheory}, these
qualitative comparisons indicate that the predictions are in reasonable
agreement with the data for all \aeta\ intervals.

To quantify the comparisons, we employ a specially derived and
previously studied $\chi^{2}$ statistic of the form~\cite{mythesis,jetsPRD}
$\chi^{2} = \sum_{i,\,j} \left( {\mathrm D}_{i}  -  {\mathrm T}_{i} \right) 
\left[ \left( {\mathrm T}_{i}/{\mathrm D}_{i} \right) {\mathrm C}_{ij} 
\left( {\mathrm T}_{j}/{\mathrm D}_{j} \right) \right]^{-1} 
\left( {\mathrm D}_{j} - {\mathrm T}_{j} \right)$, where
$\left( {\mathrm D}_{i} - {\mathrm T}_{i} \right)$ is the deviation
of the measured cross section $\left( {\mathrm D}_{i} \right)$ from the
prediction $\left( {\mathrm T}_{i} \right)$ in the $i$-th bin, 
${\mathrm C}_{ij}$ is the full covariance matrix of the 
measurement~\cite{data_on_www}, defined as
$\sum_{\alpha} \rho_{ij}^{\alpha} \sigma_{i}^{\alpha} \sigma_{j}^{\alpha}$,
where the sum runs over all sources of uncertainties, 
$\rho_{ij}$ is the correlation coefficient between the $i$-th and
$j$-th bins, and $\sigma_{i}$ is the uncertainty in the $i$-th bin.
The T/D factors are introduced to reduce the bias towards lower values
of $\chi^{2}$ originating from highly correlated systematic uncertainties 
present in ${\mathrm C}_{ij}$~\cite{jetsPRD}.
There are $90$ \peta--\et\ bins in this measurement. 

While the statistical uncertainties are not correlated in \et\ or \peta,
the systematic uncertainties are fully correlated in both variables 
except for:
(i) efficiencies for data selection, which are uncorrelated in $\eta$, 
(ii) parameterizations of jet energy resolutions and fits to the unfolding ansatz, 
which are uncorrelated in \peta, 
(iii) the hadronic response, which is partially correlated in \et\ and \peta, 
with the correlation matrix in terms of average bin energies given in 
Ref.~\cite{d0jes}.
Uncertainties in the showering correction arise dominantly from the lack of full  
agreement of the lateral shower profiles observed in the data and in the Monte 
Carlo.
The residual discrepancy is similar for all \et\ and \peta\ regions.
Consequently, the correlations of the showering correction  are large in 
\et~\cite{CentralJetsPRL} as well as in \peta.
Uncertainties due to jet energy calibration are the dominant source
of error in the cross section and range from about $12$--$20\%$ at
lowest \et\ to about $35$--$80\%$ at highest \et, getting larger with
\peta\ for a fixed \et.
They are driven by the uncertainties due to the hadronic response 
parameterization at high \et\ and due to the showering correction at 
high \et\ and, notably, at high \peta.
The second largest source of uncertainty is the jet energy resolution 
parameterization and the unfolding procedure which typically gets worse 
at low and at high \et\ and ranges from about $3$--$5\%$ at lowest \et\ 
to about $10$--$20\%$ at highest \et. 
These are followed by the uncertainties due to integrated luminosity which 
are approximately $6\%$ ($8\%)$ for the data collected with the jet filters 
with two highest (lowest) \et\ thresholds, and by the uncertainties due to 
data selection which are on the order of $1\%$ throughout the dynamic range 
of the measurement~\cite{mythesis}.

For all PDFs we have considered, Table~\ref{tab:ratios} lists the $\chi^{2}$, 
$\chi^{2}$/dof, and the corresponding probabilities for 90 degrees of freedom (dof).
We have verified that the variations of correlation coefficients within the range 
of their uncertainties give a similar ordering of the $\chi^{2}$, hence a similar
relative preference of PDFs.
The absolute values of $\chi^{2}$ and associated probabilities vary somewhat with 
variations in the correlations in \et\ and, to a much lesser extent, with variations 
of correlations in \peta.
The theoretical predictions are in good quantitative agreement with the experimental
results.
The data indicate a preference for the CTEQ4HJ, MRSTg$\uparrow$, and CTEQ4M PDFs.
The CTEQ4HJ PDF has enhanced gluon content at large $x$, favored by previous measurements 
of inclusive jet cross sections at small \peta~\cite{CentralJetsCDF,CentralJetsPRL}, 
relative to the CTEQ4M PDF.
The MRSTg$\uparrow$ PDF includes no intrinsic parton transverse momentum and therefore
has effectively increased gluon distributions at all $x$ relative to the MRST PDF.

\begin{table}[!h]
\caption{The $\chi^{2}$, $\chi^{2}$/dof, and the corresponding probabilities
         for 90 degrees of freedom for various PDFs studied.}
\begin{center}
\begin{tabular}{lccc}
{\bf PDF} & \boldmath $\chi^{2}$ &  \boldmath $\chi^{2}/\mathrm{dof}$ & {\bf Probability} \\ \hline
CTEQ3M            &  121.56  &   1.35  &    0.01   \\ 
CTEQ4M            &   92.46  &   1.03  &    0.41   \\ 
CTEQ4HJ           &   59.38  &   0.66  &    0.99   \\ 
MRST              &  113.78  &   1.26  &    0.05   \\ 
MRSTg$\downarrow$ &  155.52  &   1.73  & $<$0.01   \\
MRSTg$\uparrow$   &   85.09  &   0.95  &    0.63   \\
\end{tabular}
\end{center}
\label{tab:ratios}
\end{table}

In conclusion, we have reported a new measurement of the pseudorapidity 
and transverse-energy dependence of the inclusive jet cross section in 
proton-antiproton collisions at $\sqrt{s}=1.8$ TeV.
Our results extend significantly the kinematic reach of previous studies, 
are consistent with QCD calculations over the large dynamic range accessible 
to \dzero\ ($\aeta<3$), and indicate a preference for certain PDFs.
Once incorporated into revised modern PDFs, these measurements will
greatly improve our understanding of the structure of the proton
at large $x$ and $Q^{2}$.

\input{acknowledgement_paragraph}
\end{document}

%% file: labelfig.tex
  \ifx\LabelFigloaded\MYundefined\relax
  \else
   \fi

  \def\LabelFigloaded{\relax}

  \chardef\LabelFigCatAt\the\catcode`\@
  \catcode`\@=11

 \let\LabelFigwlog@ld\wlog
 \def\wlog#1{\relax}

 \ifx\\\MYundefined@
    \let\\\relax
 \fi

  \def\ms@g{\immediate\write16}

 \def\N@wif{\csname newif\endcsname }
 \def\Temp@ {\N@wif\ifIN@}
 \ifx\INN@\MYundefined@
    \else \let\Temp@\relax
 \fi
 \Temp@

  \def\IN@{\expandafter\INN@\expandafter}
  \long\def\INN@0#1@#2@{\long\def\NI@##1#1##2##3\ENDNI@
    {\ifx\m@rker##2\IN@false\else\IN@true\fi}%
     \expandafter\NI@#2@@#1\m@rker\ENDNI@}
  \def\m@rker{\m@@rker}
 
  \newtoks\Initialtoks@  \newtoks\Terminaltoks@
  \def\SPLIT@{\expandafter\SPLITT@\expandafter}
  \def\SPLITT@0#1@#2@{\def\TTILPS@##1#1##2@{%
     \Initialtoks@{##1}\Terminaltoks@{##2}}\expandafter\TTILPS@#2@}

 \def\Shifted@@#1#2#3{\setbox0=\hbox{#3}%
   \raise -\dp0\vbox {\kern-#2%
       \hbox {\kern#1\box0\kern-#1}%
           \kern#2}}

 \newcount\gridcount
 \newbox\auxGridbox@ \newbox\hGridbox@ \newbox\vGridbox@
 \newbox\Labelbox@ \newbox\auxLabelbox@
 \newbox\Coordinatebox@
 \newtoks\Labeltoks@
 \newdimen\Wdd@ \newdimen\Htt@

 \def\hRule@{\advance\gridcount -2%
   \vskip-.2pt\hrule\vskip-.2pt\vfil
   \llap{\smash{\raise -2.5pt
     \hbox{.\number\gridcount\kern2pt}}}%
           \vskip-.2pt\hrule\vskip-.2pt\vfil}

 \def\vRule@{\advance\gridcount 2%
   \hskip-.2pt\vrule\hskip-.2pt\hfil
   \setbox\auxGridbox@=\vbox to 0pt
      {\vskip \Htt@\vskip 2pt
        \hbox{\kern-3.5pt.\number\gridcount}\vss}%
      \wd\auxGridbox@=0pt \box\auxGridbox@
   \hskip-.2pt\vrule\hskip-.2pt\hfil}

 \def\PlaceGrid@@{\gridcount=10%
  \setbox\hGridbox@=%
    \hbox{\hbox{\hskip-.4pt\vrule
             \vbox to \Htt@{\offinterlineskip\parindent=\z@\relax
                \vskip-.4pt\hrule\vfil
                \hRule@\hRule@\hRule@\hRule@
                  \vskip-.2pt\hrule\vskip-.2pt\vfil
                \hbox to \Wdd@{\hfil}%
             \hrule\vskip-.4pt}%
         \vrule\hskip-.4pt}}%
  \gridcount=0%
  \setbox\vGridbox@=
     \hbox{\vbox{\offinterlineskip\parindent=0pt\hsize=0pt
       \vskip-.4pt\hrule%
             \hbox to \Wdd@{%
                \hskip-.4pt\vrule\hfil
                \vtop to \Htt@{\vfil}%
                     \vRule@\vRule@\vRule@\vRule@
                     \hskip-.2pt\vrule\hskip-.2pt\hfil
             \vrule\hskip-.4pt}%
         \hrule\vskip-.4pt}}%
  \wd\hGridbox@=0pt\ht\hGridbox@=0pt
  \wd\vGridbox@=0pt\ht\vGridbox@=0pt
  \hbox{\box\hGridbox@\box\vGridbox@}%
  }

 \def\LabelsGlobal{\def\LabGl@b{\global}}
 \def\LabelsLocal{\def\LabGl@b{}}
 \LabelsGlobal 

 \def\SetLabels#1\endSetLabels{%
   \LabGl@b\Labeltoks@={#1()\\}%
   }

 \LabGl@b\Labeltoks@={()\\}

 \let\PlaceGrid@\relax
 \def\ShowGrid{\let\PlaceGrid@\PlaceGrid@@}

 \def\bAdjust@@{%
     \setbox\auxLabelbox@=\hbox{\raise \dp\auxLabelbox@
            \box\auxLabelbox@}}
 \def\bAdjust@{\let\vAdjust@\bAdjust@@}

 \def\tAdjust@@{%
     \setbox\auxLabelbox@=\hbox{\raise-\ht\auxLabelbox@
            \box\auxLabelbox@}}
 \def\tAdjust@{\let\vAdjust@\tAdjust@@}

 \let\vAdjust@\relax

 \def\lAdjust@{\let\hAdjust@\rlap}
 \def\rAdjust@{\let\hAdjust@\llap}

 \let\hAdjust@\relax\let\vAdjust@\relax

 \def\FetchLabel@#1(#2)#3\\{%
     \IN@0#2@@\ifIN@
        \setbox0=\hbox{\ignorespaces#1#3\unskip}%
        \ifdim\wd0>0pt
           \ms@g{}%
           \ms@g{ !!! Bad label(s)? !!!}%
           \message{ #1(#2)#3}%
        \fi
        \def\LabelMole@##1\endFetchLabel@{%
            \IN@0()\\@##1@%
            \ifIN@\def\Temp@{\FetchLabel@##1\endFetchLabel@}%
            \else\def\Temp@{}%
            \fi
            \Temp@
           }%
     \else
       \ignorespaces#1\unskip
       \setbox\auxLabelbox@=%
         \hbox to 0pt{\hss\ignorespaces\hAdjust@
          {\ignorespaces#3\unskip}\hss}%
       \vAdjust@
       \let\hAdjust@\relax\let\vAdjust@\relax
       \AugmentLabelBox@@{#2}%
       \ht\Labelbox@=0pt\dp\Labelbox@=0pt
       \let\LabelMole@\FetchLabel@%
     \fi\LabelMole@}

 \newtoks\XYSep@ 
 \def\SetXYSeparator#1{%
     \IN@0#1@@\ifIN@\XYSep@{*}%
     \else
     \XYSep@{#1}%
     \fi
     }

 \SetXYSeparator*

 \def\AugmentLabelBox@@#1{%
     \IN@0\the\XYSep@ @#1@\ifIN@
       \SPLIT@0\the\XYSep@ @#1@%
       \setbox\Labelbox@=\hbox to 0pt{%
         \box\Labelbox@
         \Shifted@@{\the\Initialtoks@\Wdd@}%
         {\the\Terminaltoks@\Htt@}%
         {\box\auxLabelbox@}}%
     \else
         \ms@g{}%
         \ms@g{ !!! Bad insertion point. !!!}%
         \message{ (#1this point was rejected.)}%
     \fi
    }

 \def\PlaceLabels@@{\bgroup\mathsurround=0pt%
     \def\Cr@{\\}%
     \let\L\lAdjust@\let\R\rAdjust@
     \let\B\bAdjust@\let\T\tAdjust@
     \expandafter\FetchLabel@\the\Labeltoks@\endFetchLabel@
     \box\Labelbox@\egroup
     }%

 \let \PlaceLabels@\PlaceLabels@@

 \def\AffixLabels#1{\setbox\Coordinatebox@=\hbox{#1}%
      \Wdd@=\wd\Coordinatebox@ \Htt@=\ht\Coordinatebox@
      \advance\Htt@ \dp \Coordinatebox@
      \hbox{\copy\Coordinatebox@\kern-\Wdd@%
           \Shifted@@{0pt}{-\dp\Coordinatebox@}%
            {\PlaceGrid@\PlaceLabels@}%
           \kern\Wdd@}
      \let\PlaceGrid@\relax
      \LabGl@b\Labeltoks@{()\\}%
      }
 
   \let\wlog\LabelFigwlog@ld 
   \catcode`\@=\LabelFigCatAt  


%% file: definitions.tex
\newcommand{\subss}[1]{\mbox{\scriptscriptstyle \mathit{#1}}}

\newcommand{\dzero}{\mbox{D\O}}

\newcommand{\z}{\mbox{$z$}}

\newcommand{\et}{\mbox{$E_{T}$}}

\newcommand{\peta}{\mbox{$\eta$}}				
\newcommand{\aeta}{\mbox{$|\eta|$}}				
\newcommand{\ipb}{pb$^{-1}$}

\newcommand{\met}{\mbox{${\hbox{$E$\kern-0.63em\lower-.18ex\hbox{/}}}_{T}$}}
\newcommand{\metvec}{\mbox{${\hbox{$\vec{E}$\kern-0.63em\lower-.18ex\hbox{/}}}_{T}\,$}}
\newcommand{\metx}{\mbox{${\hbox{$E$\kern-0.63em\lower-.18ex\hbox{/}}}_{x}\,$}}
\newcommand{\mety}{\mbox{${\hbox{$E$\kern-0.63em\lower-.18ex\hbox{/}}}_{y}\,$}}

\newcommand{\R}{\mbox{$R_{\subss{MTE}}$}}

\newcommand{\Rmathcal}{\mbox{$\mathcal{R}$}}

\newcommand{\etal}{{\it et al.}}
\newcommand{\ppbar}{\mbox{$p\overline{p}$}}

\newcommand{\mycs}{\mbox{$d^{\,2}\sigma/(d\et d\eta)$}}

\newcommand{\T}[1]{\mbox{#1 TeV}}

\def\D0{D\O}
\def\ETmiss{{\rm {\mbox{$E\kern-0.57em\raise0.19ex\hbox{/}_{T}$}}}}

\def\simge
{\mathrel{\rlap{\raise 0.53ex \hbox{$>$}}{\lower 0.53ex \hbox{$\sim$}}}}
\def\simle
{\mathrel{\rlap{\raise 0.53ex \hbox{$<$}}{\lower 0.53ex \hbox{$\sim$}}}}

\def\ETmiss{\mbox{${\hbox{$E$\kern-0.5em\lower-.1ex\hbox{/}\kern+0.15em}}_{\rm T}$}}

\def\1800{$\sqrt{s}=1800$ GeV}
\def\630{$\sqrt{s}=630$ GeV}


%% file: list_of_authors.tex
%
\author{                                                                      
B.~Abbott,$^{56}$                                                             
A.~Abdesselam,$^{11}$                                                         
M.~Abolins,$^{49}$                                                            
V.~Abramov,$^{24}$                                                            
B.S.~Acharya,$^{16}$                                                          
D.L.~Adams,$^{58}$                                                            
M.~Adams,$^{36}$                                                              
G.A.~Alves,$^{2}$                                                             
N.~Amos,$^{48}$                                                               
E.W.~Anderson,$^{41}$
% ...
\\                                                                            
\vskip 0.30cm                                                                 
\centerline{(D\O\ Collaboration)}                                             
\vskip 0.30cm                                                                 
}                                                                             
\address{                                                                     
\centerline{$^{1}$Universidad de Buenos Aires, Buenos Aires, Argentina}       
\centerline{$^{2}$LAFEX, Centro Brasileiro de Pesquisas F{\'\i}sicas,         
                  Rio de Janeiro, Brazil}                                     
\centerline{$^{3}$Universidade do Estado do Rio de Janeiro,                   
                  Rio de Janeiro, Brazil}                                     
\centerline{$^{4}$Institute of High Energy Physics, Beijing,                  
                  People's Republic of China}                                 
\centerline{$^{5}$Universidad de los Andes, Bogot\'{a}, Colombia}             
\centerline{$^{6}$Charles University, Prague, Czech Republic}                 
\centerline{$^{7}$Institute of Physics, Academy of Sciences, Prague,          
                  Czech Republic}                                             
\centerline{$^{8}$Universidad San Francisco de Quito, Quito, Ecuador}         
\centerline{$^{9}$Institut des Sciences Nucl\'eaires, IN2P3-CNRS,             
                  Universite de Grenoble 1, Grenoble, France}                 
\centerline{$^{10}$CPPM, IN2P3-CNRS, Universit\'e de la M\'editerran\'ee,     
                  Marseille, France}                                          
\centerline{$^{11}$Laboratoire de l'Acc\'el\'erateur Lin\'eaire,              
                  IN2P3-CNRS, Orsay, France}                                  
\centerline{$^{12}$LPNHE, Universit\'es Paris VI and VII, IN2P3-CNRS,         
                  Paris, France}                                              
\centerline{$^{13}$DAPNIA/Service de Physique des Particules, CEA, Saclay,    
                  France}                                                     
\centerline{$^{14}$Panjab University, Chandigarh, India}                      
\centerline{$^{15}$Delhi University, Delhi, India}                            
\centerline{$^{16}$Tata Institute of Fundamental Research, Mumbai, India}     
\centerline{$^{17}$Seoul National University, Seoul, Korea}                   
\centerline{$^{18}$CINVESTAV, Mexico City, Mexico}                            
\centerline{$^{19}$FOM-Institute NIKHEF and University of                     
                  Amsterdam/NIKHEF, Amsterdam, The Netherlands}               
\centerline{$^{20}$University of Nijmegen/NIKHEF, Nijmegen, The               
                  Netherlands}                                                
\centerline{$^{21}$Institute of Nuclear Physics, Krak\'ow, Poland}            
\centerline{$^{22}$Institute for Theoretical and Experimental Physics,        
                   Moscow, Russia}                                            
\centerline{$^{23}$Moscow State University, Moscow, Russia}                   
\centerline{$^{24}$Institute for High Energy Physics, Protvino, Russia}       
\centerline{$^{25}$Lancaster University, Lancaster, United Kingdom}           
\centerline{$^{26}$Imperial College, London, United Kingdom}                  
\centerline{$^{27}$University of Arizona, Tucson, Arizona 85721}              
\centerline{$^{28}$Lawrence Berkeley National Laboratory and University of    
                  California, Berkeley, California 94720}                     
\centerline{$^{29}$University of California, Davis, California 95616}         
\centerline{$^{30}$California State University, Fresno, California 93740}     
\centerline{$^{31}$University of California, Irvine, California 92697}        
\centerline{$^{32}$University of California, Riverside, California 92521}     
\centerline{$^{33}$Florida State University, Tallahassee, Florida 32306}      
\centerline{$^{34}$University of Hawaii, Honolulu, Hawaii 96822}              
\centerline{$^{35}$Fermi National Accelerator Laboratory, Batavia,            
                   Illinois 60510}                                            
\centerline{$^{36}$University of Illinois at Chicago, Chicago,                
                   Illinois 60607}                                            
\centerline{$^{37}$Northern Illinois University, DeKalb, Illinois 60115}      
\centerline{$^{38}$Northwestern University, Evanston, Illinois 60208}         
\centerline{$^{39}$Indiana University, Bloomington, Indiana 47405}            
\centerline{$^{40}$University of Notre Dame, Notre Dame, Indiana 46556}       
\centerline{$^{41}$Iowa State University, Ames, Iowa 50011}                   
\centerline{$^{42}$University of Kansas, Lawrence, Kansas 66045}              
\centerline{$^{43}$Kansas State University, Manhattan, Kansas 66506}          
\centerline{$^{44}$Louisiana Tech University, Ruston, Louisiana 71272}        
\centerline{$^{45}$University of Maryland, College Park, Maryland 20742}      
\centerline{$^{46}$Boston University, Boston, Massachusetts 02215}            
\centerline{$^{47}$Northeastern University, Boston, Massachusetts 02115}      
\centerline{$^{48}$University of Michigan, Ann Arbor, Michigan 48109}         
\centerline{$^{49}$Michigan State University, East Lansing, Michigan 48824}   
\centerline{$^{50}$University of Nebraska, Lincoln, Nebraska 68588}           
\centerline{$^{51}$Columbia University, New York, New York 10027}             
\centerline{$^{52}$University of Rochester, Rochester, New York 14627}        
\centerline{$^{53}$State University of New York, Stony Brook,                 
                   New York 11794}                                            
\centerline{$^{54}$Brookhaven National Laboratory, Upton, New York 11973}     
\centerline{$^{55}$Langston University, Langston, Oklahoma 73050}             
\centerline{$^{56}$University of Oklahoma, Norman, Oklahoma 73019}            
\centerline{$^{57}$Brown University, Providence, Rhode Island 02912}          
\centerline{$^{58}$University of Texas, Arlington, Texas 76019}               
\centerline{$^{59}$Texas A\&M University, College Station, Texas 77843}       
\centerline{$^{60}$Rice University, Houston, Texas 77005}                     
\centerline{$^{61}$University of Virginia, Charlottesville, Virginia 22901}   
\centerline{$^{62}$University of Washington, Seattle, Washington 98195}       
}                                                                             

%% file: acknowledgement_paragraph.tex
%
We thank the staffs at Fermilab and collaborating institutions, 
and acknowledge support from the 
Department of Energy and National Science Foundation (USA),  
Commissariat  \` a L'Energie Atomique and 
CNRS/Institut National de Physique Nucl\'eaire et 
de Physique des Particules (France), 
Ministry for Science and Technology and Ministry for Atomic 
   Energy (Russia),
CAPES and CNPq (Brazil),
Departments of Atomic Energy and Science and Education (India),
Colciencias (Colombia),
CONACyT (Mexico),
Ministry of Education and KOSEF (Korea),
CONICET and UBACyT (Argentina),
The Foundation for Fundamental Research on Matter (The Netherlands),
PPARC (United Kingdom),
A.P. Sloan Foundation,
and the A. von Humboldt Foundation.

%% file: lb_prp.bbl
\begin{thebibliography}{99}
\bibitem{data_on_www}
	Data for all \aeta\ regions and covariance matrices are available from 
	the AIP E-PAPS service.
\end{thebibliography}
